\documentclass[reprint, longbibliography]{revtex4-1}

\usepackage{mathtools}
\usepackage{amsmath}
\usepackage{amssymb}
\usepackage{color}

\usepackage[normalem]{ulem}

\begin{document}

\title{Extended self-similarity in moment-generating-functions in wall-bounded turbulence at high Reynolds number}

\author{ X. I. A. Yang$^1$, C. Meneveau$^1$, I. Marusic$^2$ and L. Biferale$^3$\\
$^1$ Department of Mechanical Engineering and Center for Environmental and Applied Fluid Mechanics, The Johns Hopkins University, 3400 N. Charles Street, Baltimore, Maryland 21218, USA \\
$^2$ Department of Mechanical Engineering, University of Melbourne, Victoria 3010, Australia\\
$^3$Dipartimento di Fisica and INFN, Universita di Roma ``Tor Vergata,'' Via della Ricerca Scientifica 1, 00133 Roma, Italy
}

%
%

\begin{abstract}
In wall-bounded turbulence, the moment generating functions (MGFs) of the streamwise velocity fluctuations $\left<\exp(qu_z^+)\right>$ develop power-law scaling as a function of the wall normal distance $z/\delta$. Here $u$ is the streamwise velocity fluctuation, $+$ indicates normalization in wall units (averaged friction velocity), $z$ is the distance from the wall, $q$ is an independent variable and $\delta$ is the boundary layer thickness. 
Previous work has shown that this power-law scaling exists in the log-region {\small $3Re_\tau^{0.5}\lesssim z^+$, $z\lesssim 0.15\delta$}， where $Re_\tau$ is the friction velocity-based Reynolds numbers. 
Here we present empirical evidence that this self-similar scaling can be extended, including bulk and viscosity-affected regions $30<z^+$, $z<\delta$, provided the data are interpreted with the Extended-Self-Similarity (ESS), i.e. self-scaling of the MGFs as a function of one reference value, $q_o$. 
ESS also improves the scaling properties, leading to more precise measurements of the scaling exponents. 
The analysis is based on hot-wire measurements from boundary layers at $Re_\tau$ ranging from $2700$ to $13000$ from the Melbourne High-Reynolds-Number-Turbulent-Boundary-Layer-Wind-Tunnel. 
Furthermore, we investigate the scalings of the filtered, large-scale velocity fluctuations $u^L_z$ and of the remaining small-scale component, $u^S_z=u_z-u^L_z$. The scaling of $u^L_z$ falls within the conventionally defined log region and depends on a scale that is proportional to {\small $l^+\sim Re_\tau^{1/2}$}; the scaling of $u^{S}_z$ extends over a much wider range from $z^+\approx 30$ to $z\approx 0.5\delta$.
Last, we present a theoretical construction of two multiplicative processes for $u^L_z$ and $u^S_z$ that reproduce the empirical findings concerning the scalings properties as functions of $z^+$ and in the ESS sense.

\textbf{Postprint version of the article published on Physical Review Fluids vol 1, 044405 (2016). DOI: 10.1103/PhysRevFluids.1.044405}
\end{abstract}
\maketitle

\section{Introduction}

Turbulent boundary layers have been one of the centerpieces of turbulence research for many decades \citep{marusic2010wall, jimenez2011cascades, smits2011high}. A robust feature of wall-bounded flows is the logarithmic scaling of the mean velocity in the log region $U/u_\tau=1/\kappa \ln(z^+)+B$ \citep{prandtl1925report, hultmark2012turbulent, marusic2013logarithmic, lee2015direct}, where $U$ is the mean velocity, $u_\tau$ is the friction velocity, $z$ is the wall normal distance, the superscript $+$ indicates normalization by wall units. Recently, there has been growing empirical evidence for a logarithmic scaling in the 
variance of the streamwise velocity fluctuations, $\left<(u^+_z)^2\right>=A_1\ln(\delta/z)+B_1$ \citep{marusic2003streamwise, kunkel2006study, marusic2013logarithmic, stevens2014large, lee2015direct}, where $A_1\approx 1.26$ is the Townsend-Perry constant, $B_1$ is yet another constant, and $\delta$ is an outer length scale. 
Motivated by calculations based on the Townsend attached eddy hypothesis \citep{townsend1976structure}, in Ref. \cite{meneveau2013generalized}, Meneveau \& Marusic observed logarithmic scalings in {\small $\left<(u_z^{+})^{2p}\right> ^{1/p}$}; logarithmic scalings are also found in the longitudinal structure functions {\small $\left<(u^+_z(x+r)-u^+_z(x))^{2p}\right>^{1/p}$}, where $r$ is taken along the streamwise direction at a fixed distance from the wall \citep{de2015scaling}.

To study the flow physics from a new perspective, it was proposed in Ref. \cite{yang2016moment} to shift the attention to the MGFs of the single-point streamwise velocity fluctuations: $W(q,z/\delta)= \left<\exp(qu_z^+)\right>$ where $\left<\cdot\right>$ indicates ensemble averaging and $q$ is an independent variable that serves as a ``dial'' to emphasize different turbulent fluctuation intensities and signs. In Ref. \cite{yang2016moment} empirical evidence of a power-law behavior in the log-region for the MGFs was presented (see also Fig.\ref{fig:expqu-ESS}a in this paper), leading to the introduction of the scaling exponents $\tau(q)$ according to
\begin{equation}
\small
W(q,z/\delta)= \left<\exp(qu_z^+)\right> \sim (\delta/z)^{\tau(q)}.
\label{eq:powlaw}
\end{equation}

From the perspective of modeling, the logarithmic scalings in $U$, $\left<(u_z^+)^{2p}\right>^{1/p}$, $\left<(u^+(x+r)-u^+(x))^{2p}\right>^{1/p}$, and the power-law scaling in $\left<\exp(qu_z^+)\right>$ evidence the presence of self-similar, space-filling, wall-attached eddies in the log region \citep{perry1982mechanism, perry1994wall, marusic1995wall, woodcock2015statistical}. Modeling the wall turbulence as collections of self-similar, space-filling, wall-attached eddies is the basic idea of the attached eddy model.
As has been reviewed, the attached eddy model is quite useful in providing (non-trivial) estimates on the scaling behaviors of flow statistics in the log region; and since Townsend \cite{townsend1976structure}, this model has been providing guidance to studies of the near-wall flow physics \citep{bullock1978structural, de2016uniform, yang2016moment, del2006self, klewicki2014self}. In fact, many results in this work can be understood in the framework of the attached eddy hypothesis. 

Besides the attached eddy model, in this work, we use the concept of Extended-Self-Similarity (ESS) to interpret the high Reynolds number boundary layer data. The ESS concept was developed originally for isotropic homogeneous turbulence, see, e.g. Refs. \cite{benzi1993extended, benzi1995scaling, benzi1996generalized}. 
The basic idea of ESS is to define self-similarity and statistical scaling as function of certain reference scaling. In early works on ESS in homogeneous isotropic turbulence, the specific statistical scaling is the power-law scaling in the various moments of two-point velocity increments. Using ESS, the scaling ranges were seen to extend beyond the inertial range. With the extended scaling region, the scaling exponents can be measured with higher accuracy \cite{arneodo1996structure}. Beside isotropic homogeneous turbulence, applications of ESS in anisotropic turbulence and hydromagnetic turbulence can be found in Refs.  \cite{carbone1995experimental,benzi1996extended,carbone1996evidences,benzi1996extended,
grossmann1997application}. Several explanations have been provided about the physics underlying ESS \cite{benzi1996generalized,benzi1995intermittent,meneveau1996transition}. There is general agreement that ESS helps to reduce the deviations from inertial-range scaling arising from finite-size and viscous effects at the edges of the scaling range. 

The context of this work is wall-bounded flows. Using ESS, we express the scaling of $W(q,z/\delta)$ with respect to $W(q_o,z/\delta)$ ($q_o$ fixed):
\begin{equation}
\small
W(q,z/\delta)=W(q_o,z/\delta)^{\xi(q,q_o)},
\label{eq:ESS}
\end{equation}
instead of the scaling of $W(q,z/\delta)$ with respect to $z$ as is usually done.
Eq. \ref{eq:ESS} is a quite general definition for scaling. It trivially includes the definition in Eq. \ref{eq:powlaw}, for which case $\xi(q,q_o) = \tau(q)/\tau(q_o)$. 
The idea behind the definition in Eq. \ref{eq:ESS} is as follows: for wall-bounded flows, the underlying physics does not permit perfect self-similarity in wall eddies as a function of their sizes because of the bulk flow effects (for $z\sim O(\delta)$) or viscous, dissipation effects (for $z^+\sim O(1)$); but because those effects exist in both in $W(q,z/\delta)$ and $W(q_o,z/\delta)$, by defining scalings according to Eq. \ref{eq:ESS}, self-similarity can be extended; from this extended self-similarity, {\it suitable} physics-based models can be developed.  

In this work, we show that, with the scaling defined in Eq. \ref{eq:ESS}, self-similarity in wall-turbulence extends beyond the log region, to the region $30<z^+$, $z<\delta$, with high-quality scalings. 
With the extended high-quality scaling, the quantity $\tau(q)/\tau(q_o)$ can be determined with higher accuracy than using the scalings defined in Eq. \ref{eq:powlaw}. 
Furthermore, we decompose the fluctuating velocity into a large-scale signal $u^L_z$ and a small-scale signal $u^S_z$. We present evidence that the breaking of the self-similarity (power-law of the MGFs) is because of a change in the statistical properties of $u^L_z$ at $z^+\sim Re_\tau^{0.5}$, whereas $u_z^S$ retains more universal scaling closer to the wall. 
Last, to understand the  physical processes underlying the statistical behaviors of $u_z^L$, $u_z^S$, we propose a model based on two random additive processes for $u^S_z$ and $u^L_z$. In this way, we can explain the existence of ESS in the extended region $30<z^+$, $z<\delta$, the existence of conventional scaling Eq. \ref{eq:powlaw} in the log region, and the failure of the conventional scaling beyond the log region.

\section{Attached eddy models}\label{sect:AttEddy}

Before detailed discussion of ESS, we briefly review a recently develop formalism of the attached eddy model, the hierarchical-random-additive-process \citep{meneveau2013generalized, yang2016moment}. The investigation of ESS in this work is motivated by predictions from this attached eddy formalism.
For simplicity, unless indicated otherwise, wall units are used for normalization of the velocity, and we drop the superscript $+$ hereafter. 

A direct consequence of space-filling, self-similar eddies in the log region is the eddy population density being inversely proportional to the wall normal distance, i.e., $P(z)\sim 1/z$. 
Knowing the eddy population density, and modeling the velocity fluctuation at a generic point in the flow field to be a consequence of superpositions of the attached eddy induced velocities, we can write:
\begin{equation}
\small
u_z=\sum_{i=1}^{N_z}a_i,~~~N_z=\int_{z}^\delta P(z^\prime)dz^\prime\sim\log(\delta/z).
\label{eq:hrap}
\end{equation} 
$a_i$ are identically, independently distributed random variables (i.i.d.). Each $a_i$ represents the effect of one attached eddy (self-similarity of the attached eddies leads to the property of i.i.d. of $a_i$). The number of the random additives, $N_z$, is obtained by integrating the eddy population density from $z$, the height of the point of interest, to the boundary layer height.
Taking exponential, ensemble averaging of Eq. \ref{eq:hrap} leads to the power-law scaling of the MGFs
\begin{equation}
\small
\left<\exp(qu_z)\right> = \prod_{i=1}^{N_z}\left< \exp(q a_i) \right> \sim (\delta/z)^{\tau(q)}.
\label{eq:expqu}
\end{equation}
The symbol `$\sim$' here signifies `proportional to'.
For the equality, we have used independency among $a_i$'s and for the second the property of statistical identicality.
Provided the random additive $a_i$ is Gaussian, we have: $\tau(q)\sim q^2$.
Central moments can be directly computed from the MGFs.
For example, $\left<u^2\right>=\left.d^2W(q,z/\delta)/dq^2\right|_{q=0}$.
As was shown in Ref. \cite{yang2016moment}, the Towsend-Perry constant can then be calculated accordingly: $A_1=\left.d^2\tau/dq^2\right|_{q=0}$ and measurements yielded $A_1\approx 1.26$ \cite{yang2016moment}. Near $q=0$, the behavior of $\tau(q)$ is well approximated by $\tau(q)\sim q^2$, and the data analysis  yields
\cite{yang2016moment}:
\begin{equation}
\small
\tau(q) = 0.63q^2.
\label{eq:tauq-app}
\end{equation}

Let us now anticipate the key observation of this work. 
Scaling of the MGFs requires all additives in Eq. \ref{eq:hrap} to be i.i.d. variables, i.e. wall attached-eddies being statistically independent and identical. 
We relax one of the requirements --the requirement of $a_i$'s to be statistically identical, but we maintain the request of $a_i$'s being independently distributed, i.e. we still assume noninteracting wall eddies, but we allow the eddies' characteristic velocities to depend on distance to the wall.
For example, let us take each additive being Gaussian with a scale-dependent variance $\sigma_i$.
We then have for Eq. \ref{eq:expqu}:
\begin{equation}
\small
\left<\exp(qu_z)\right> \sim \prod_{i=1}^{N_z}\left< \exp(q a_i) \right> \sim \exp\left(\frac{q^2}{2}\sum_{i=1}^{N_z} \sigma_i\right).
\label{eq:expqu2}
\end{equation}
Provided $\sigma_i$ is $i$-independent, Eq. \ref{eq:expqu2} simplifies to Eq. \ref{eq:expqu}. If $\sigma_i$ is $i$-dependent, scaling of the MGFs in the usual power-law sense does not survive.
However, ESS still holds with a $z$-independent $\xi$ 
\begin{equation}
\small
\xi(q,q_o)=\frac{\log\left<\exp(qu_z)\right>}{\log\left<\exp(q_ou_z)\right>} =\frac{q^2}{q_o^2} 
\label{eq:ess2}
\end{equation}
for any couple of $q$, $q_o$. 

Eqs. \ref{eq:expqu2}, \ref{eq:ess2} capture the basic physics behind the ESS scalings in the context of wall-bounded flows. Self-similarity holds only in the log region, where the only characteristic velocity scale is the friction velocity and the only length scale is the distance from the wall. This characteristic velocity scale is the basis of $a_i$ being assumed to be i.i.d. and the characteristic length scale is the basis of Eq. \ref{eq:hrap} with eddies being space-filling and wall-attached. Beyond the log region, it is expected that friction velocity ceases to be the only relevant velocity scale. Thus we allow the $a_i$ distribution to vary and not be independent on $z$. The existence of the ESS scalings suggests that the HRAP formalism can be extended beyond the log region by allowing the characteristic velocity to be scale-dependent.

\section{Extended Self-similarity in Wall-Bounded Flows} \label{sect:ESS}

In this section we present empirical evidence of ESS in wall-bounded flows at high Reynolds number. 
Hot-wire measurements of the streamwise velocity taken from the Melbourne High-Reynolds-Number-Boundary-Layer-Wind-Tunnel (HRNBLWT) from a boundary layer at $Re_\tau=13000$ are analyzed [with $U_\infty=20 (ms^-1)$, $u_\tau=0.639 (ms^-1)$ and $\delta=0.319(m)$, see Refs. \cite{hutchins2009hot, marusic2015evolution} for details of the dataset]. A convergence analysis of this dataset is conducted in Ref. \cite{yang2016moment}. 

\begin{figure} 
\centering
\includegraphics[height=2.00in]{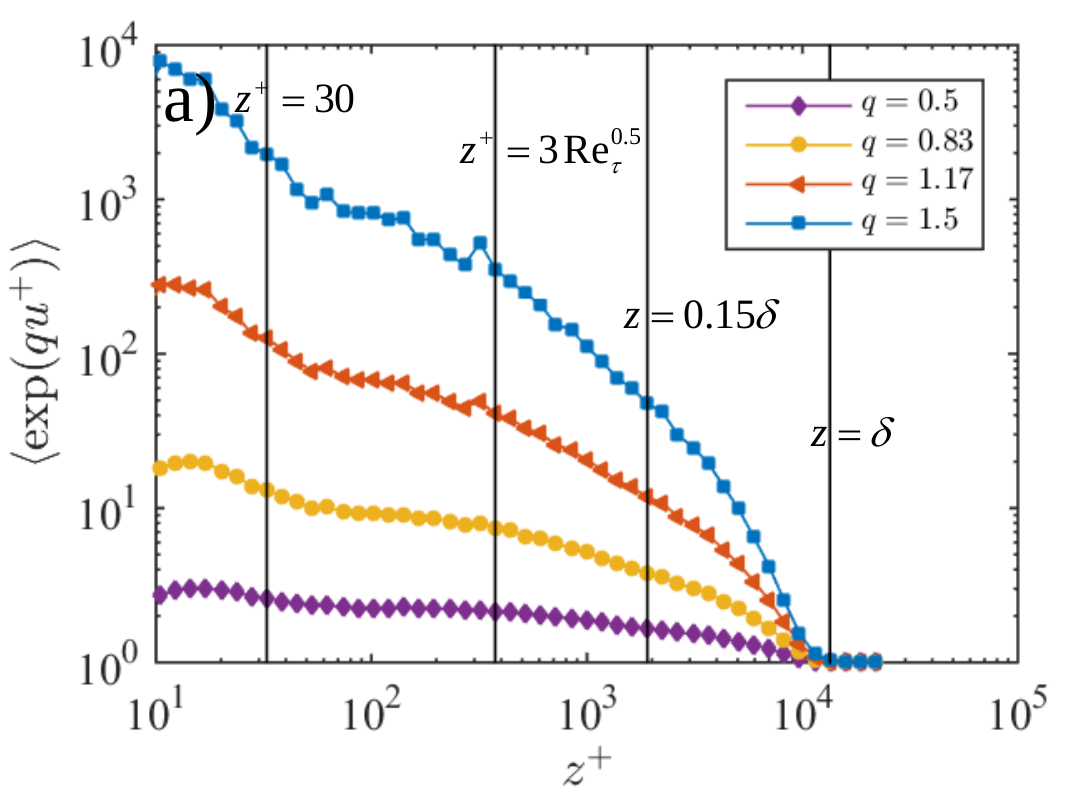}\\
\includegraphics[height=2.00in]{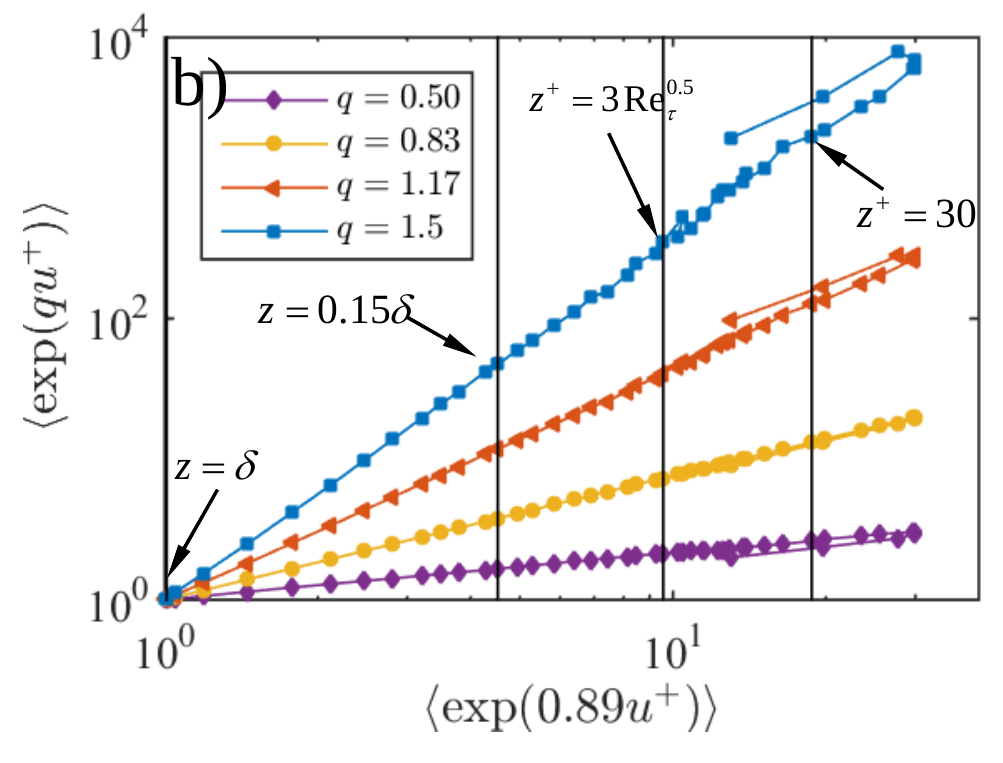}
\caption{(a) log-log plot of $\left<\exp(qu^+)\right>$ against $z^+$ for $q=0.5$, $0.83$, $1.17$, $1.5$. $z^+=30$, $3Re_\tau^{0.5}$, $z=0.15\delta$, $\delta$ are indicated in the figure using vertical lines. (b) log-log plot of $\left<\exp(qu^+)\right>$ against $\left<\exp(0.89u^+)\right>$ for $q=0.5, 0.83, 1.17, 1.5$. $z^+=30$, $3Re^{0.5}$, and $z/\delta=0.15$, $1$ are indicated for $q=1.5$. The folding back of the line corresponds to the near wall region, $z^+ \le 10$. }
\label{fig:expqu-ESS}
\end{figure}

The measured $\left<\exp(qu_z)\right>$ as functions of the wall normal distance $z^+$ are plotted on a log-log scale in Fig. \ref{fig:expqu-ESS} (a) for representative positive $q$ values, $q=0.5$, $0.83$, $1.17$, $1.5$ (see Ref. \cite{yang2016moment} for detailed discussion on the scaling of $W(q,z/\delta)$). As already anticipated, power-law scalings are observed only in the log region
$3Re_\tau^{0.5}<z^+$, $z<0.15\delta$ (see Ref. \cite{marusic2013logarithmic} for a detailed discussion on the extent of the log region in wall turbulence). 
In this section we mainly focus on $W(q,z/\delta)$ with positive $q$ values. Because $\left<\exp(qu_z)\right>$ emphasizes velocity fluctuations of the same sign as $q$, positive $q$ values emphasize high positive velocity fluctuations.
Taking positive $q$ values, we emphasize those fluid configurations which are influenced primarily by ``sweep'' motions 
\citep{kline1967structure, cantwell1981organized, katul2006relative}. 
A brief discussion on the ``ejection'' motions, corresponding to negative $q$ values, can be found at the end of this section.
In Fig. \ref{fig:expqu-ESS}(b), the measured $W(q,z/\delta)$ is plotted against $W(q_o=0.89;z/\delta)$ for positive $q$ values. The choice of $q_o$ is arbitrary because ESS requires no specific choice of $q_o$.
Here we fixed the reference scaling such that, $\tau(q_o) = 0.5$.
As is evident in the figure, extended self-similarity is observed in $30<z^+$, $z<\delta$, i.e. in a region that is significantly more extended than the log region.
Let us stress that $\tau(q)$ cannot be directly measured from the ESS scalings and only $\xi(q,q_o)= \tau(q)/\tau(q_o)$ is accessible. To recover the value of the absolute exponent $\tau(q)$, we invert the previous relation using the value of $\tau(q_o)$ obtained form the power-law scaling. 
Already on a qualitative basis, one can see that by comparing the scalings in Fig. \ref{fig:expqu-ESS}(a)-(b), the ESS scalings are of higher quality and can be observed in a more extended region. 

\begin{figure} 
\centering
\includegraphics[height=2.00in]{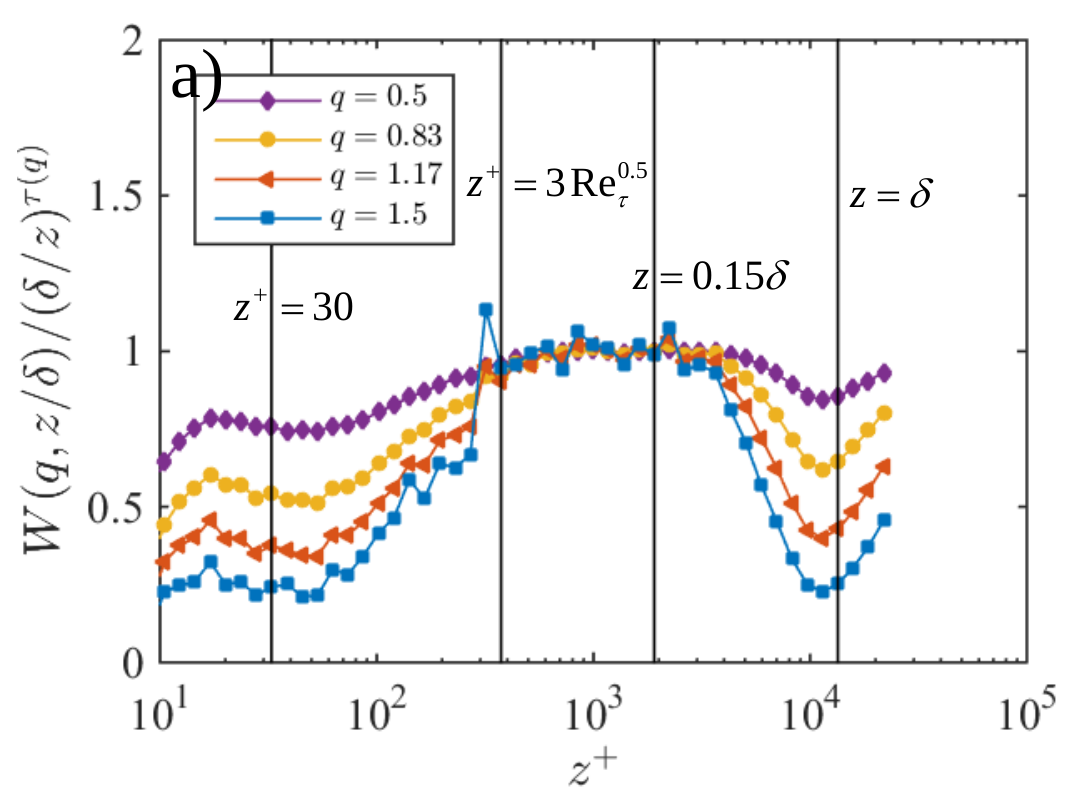}\\
\includegraphics[height=2.00in]{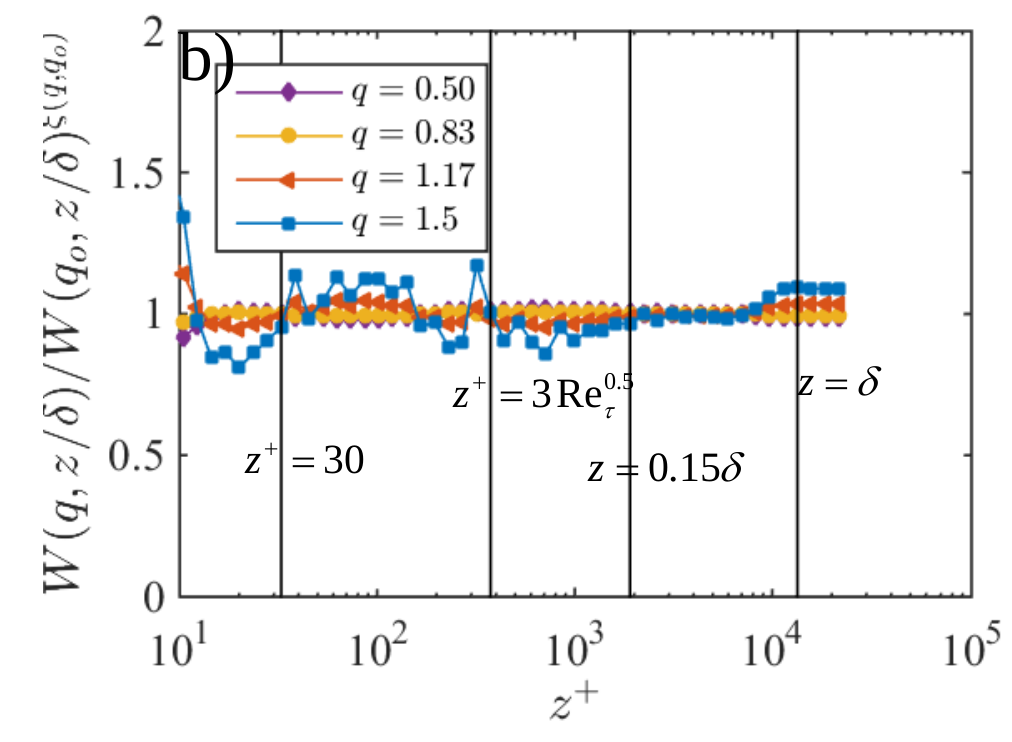}
\caption{(a) Compensated scaling, $W(q,z/\delta)/(\delta/z)^{\tau(q)}$. 
(b) Compensated ESS scaling, $W(q,z/\delta)/W(q_o,z/\delta)^{\xi(q,q_o)}$ plotted against $z^+$. 
The plots are vertically shifted to collapse at $1$. }
\label{fig:cmpn}
\end{figure}

To quantify the quality of the scalings, we plot in Fig. \ref{fig:cmpn} (a) the MGFs compensated with the regular power-law scalings $W(q,z/\delta)/(\delta/z)^{\tau(q)}$ and in (b) MGFs compensated with the ESS scalings  $W(q,z/\delta)/W(q_o,z/\delta)^{\xi(q,q_o)}$. The ESS scaling extends to the near-wall region, $30 < z^+ < 3 Re_\tau^{1/2}$, and to the bulk, $0.15 \delta < z^+ < \delta$ (deviations from unit do not exceeding $10 \%$ for the largest moment, $q=1.5$). 
On the other hand, the power-law scalings of the MGFs are observed only in the log region, above $z^+=3Re^{0.5}_\tau$ and below $z=0.15\delta$. 
 
With self-similar wall-attached eddies, the compensated scaling $W(q,z/\delta)/(\delta/z)^{\tau(q)}$ develops a plateau. As can be seen in Fig. \ref{fig:cmpn} (a), this plateau region coincides with the log region. 
Beyond the log region, the observed pleteau in $W(q,z/\delta)/W(q_o,z/\delta)^{\xi(q,q_o)}$ suggests that while the eddies' characteristic velocity scale is dependent on $z$ outside of the log region, the eddies' organization is not inconsistent with the attached eddy model (or the HRAP formalism).

\begin{figure} 
\centering
\includegraphics[height=2.00in]{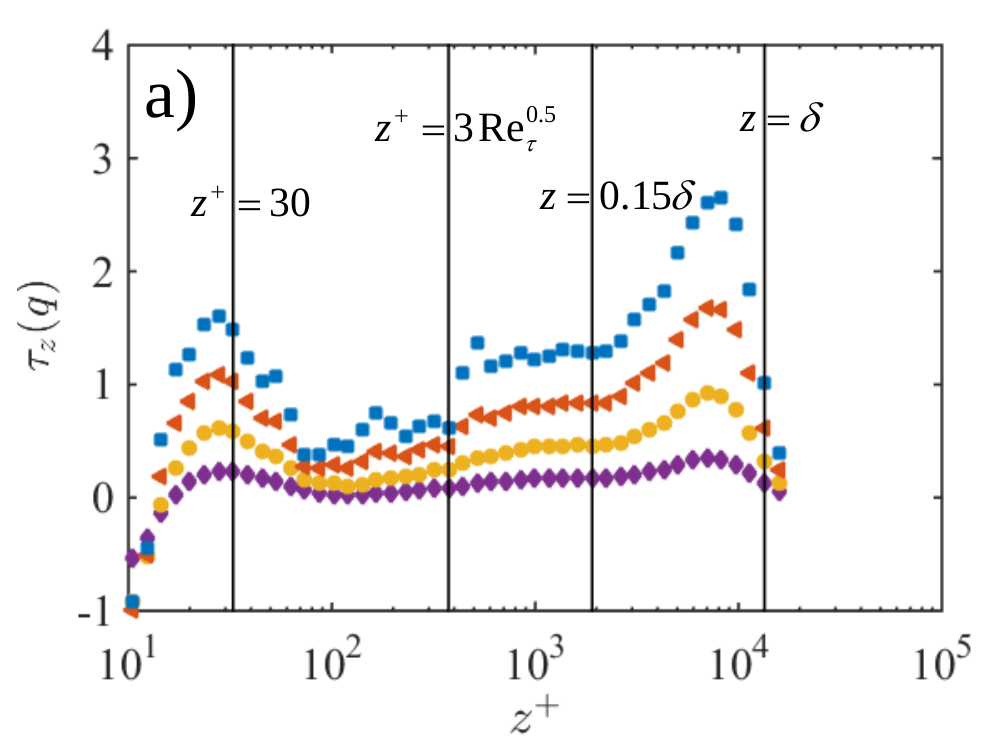}\\
\includegraphics[height=2.00in]{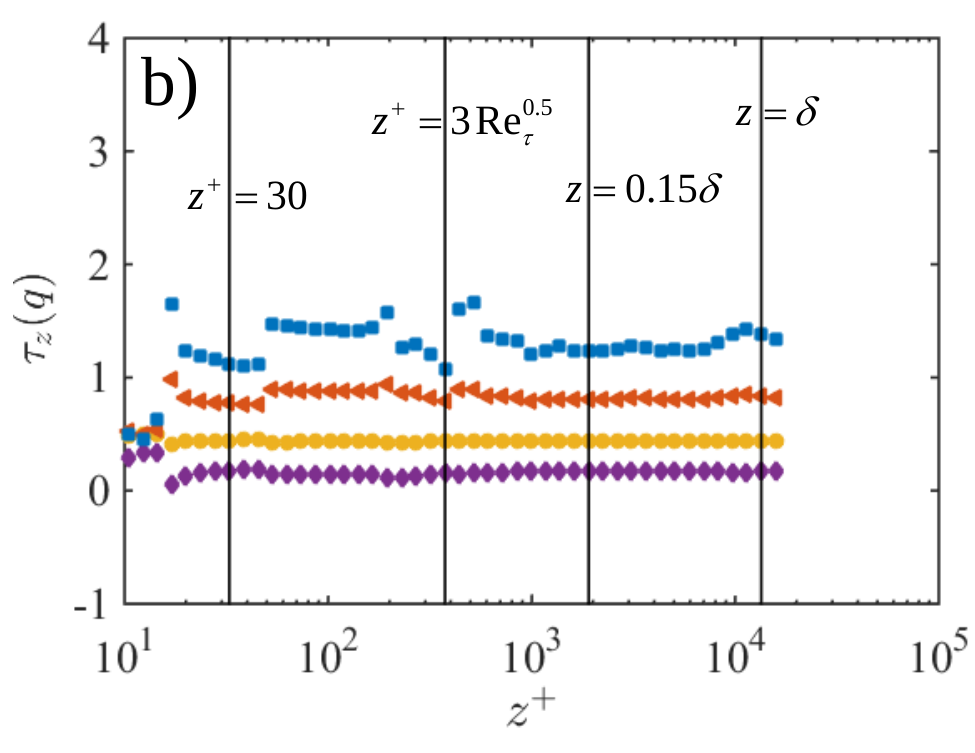}
\caption{(a) Local scaling exponent of $W(q,z/\delta)$ (Eq. \ref{eq:tqzW}). (b) Point-to-point scaling exponent of $W(q,z/\delta)$ from ESS and plotted against the wall normal distance (Eq. \ref{eq:tqzESS}). The fitted local scaling exponents locally averaged among 5 consecutive points. }
\label{fig:slp}
\end{figure}

In Fig. \ref{fig:slp} (a) and (b), we assess self-similarity {\it scale-by-scale} using
the 
local slopes of the curves shown in Fig. \ref{fig:expqu-ESS} as a function 
of the wall normal distance $z^+$. 
For the power-law scaling of MGFs (shown in Fig. 3(a)): 
\begin{equation}
\small
\tau(q,z)=\frac{d\log(W(q,z/\delta))}{d\log(z/\delta)}
\label{eq:tqzW}
\end{equation}
and for the ESS (Fig. 3(b)):
\begin{equation}
\small
\tau(q,z)=\tau(q_o)\cdot\frac{d\log(W(q,z/\delta))}{d\log(W(q_o,z/\delta))}=\tau(q_o)\cdot\xi(q,q_o,z).
\label{eq:tqzESS}
\end{equation}
Again, a significantly more extended scaling region is found for the ESS scalings. 

\begin{figure} 
\centering
\includegraphics[height=2.00in]{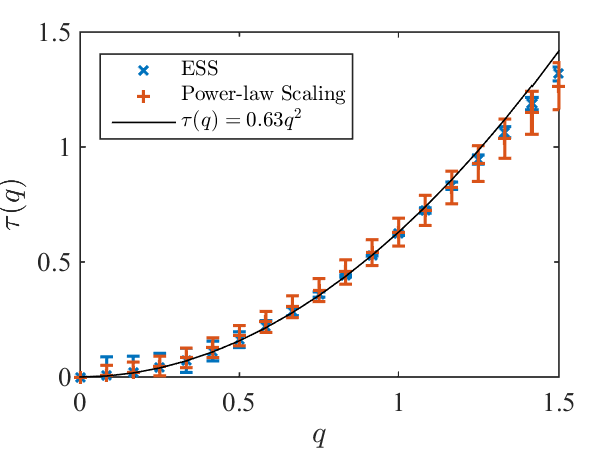}
\caption{Fitted scaling exponents of $W(q,z/\delta)$ (symbols). The power law fit ($+$) is conducted in the log region ($3Re_\tau^{0.5}<z^+$, $z<0.15\delta$). ESS fit ($\times$) is conducted in the region $30<z^+$, $z<\delta$. The solid line corresponds a best fit near $q=0$, $0.63q^2$. 
}
\label{fig:tauq}
\end{figure}

In Fig. \ref{fig:tauq}, we summarize the results for the scaling exponents for positive $q$ values.
Direct power-law fitting of $W(q,z/\delta)$ against $z/\delta$ is restricted to the log region; the ESS fit of $W(q,z/\delta)$ against $W(0.89;z/\delta)$ can be conducted in $30<z^+$, $z<\delta$. Error bars are estimated by considering a linear fit in log-log coordinate, $y \approx ax+b$. We define the uncertainty in the fitted slope, $a$, as follows: $\Delta a = 3\text{std}(y-ax-b)/[a(\max(x)-\min(x))]$. Where $\text{std}(y-ax-b)$ is the standard deviation of the residual, and  $a(\max(x)-\min(x))$ is the expected change of $y$ in the fitted range. The uncertainty in the fitted value of  $a$ is taken to be three times the ratio of those two quantities. 
Up to $q=1$, the measured scaling exponent follow the Gaussian approximation closely.
ESS clearly helps in reducing the uncertainty in measuring the scaling exponents. 

\begin{figure} 
\centering
\includegraphics[height=2.00in]{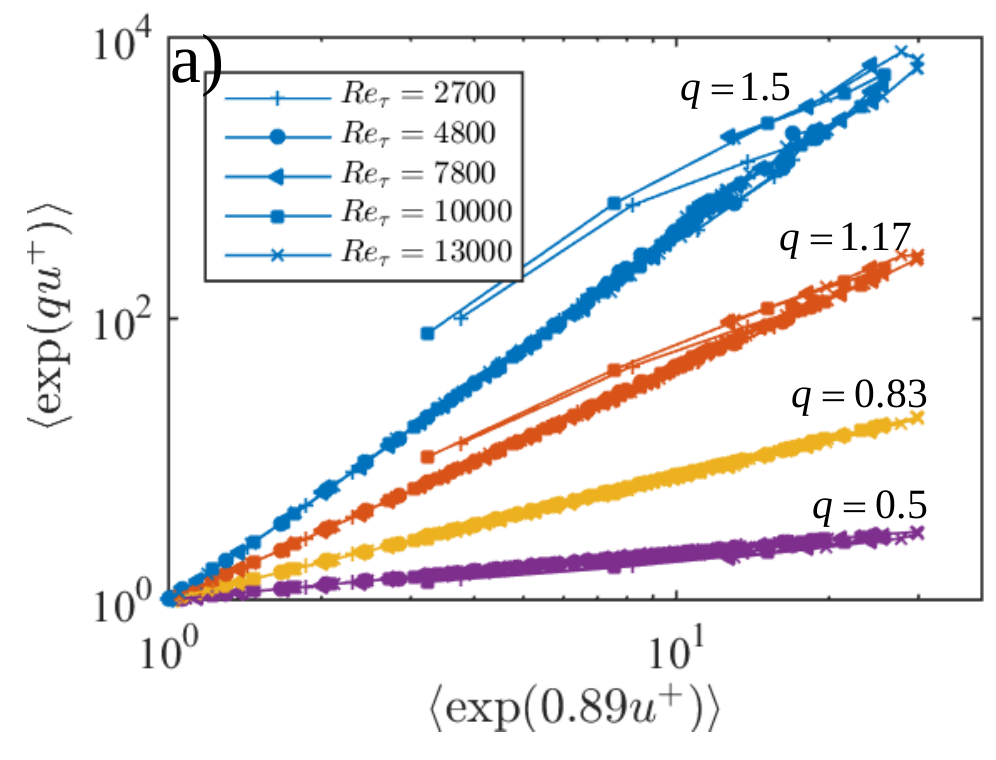}
\includegraphics[height=2.00in]{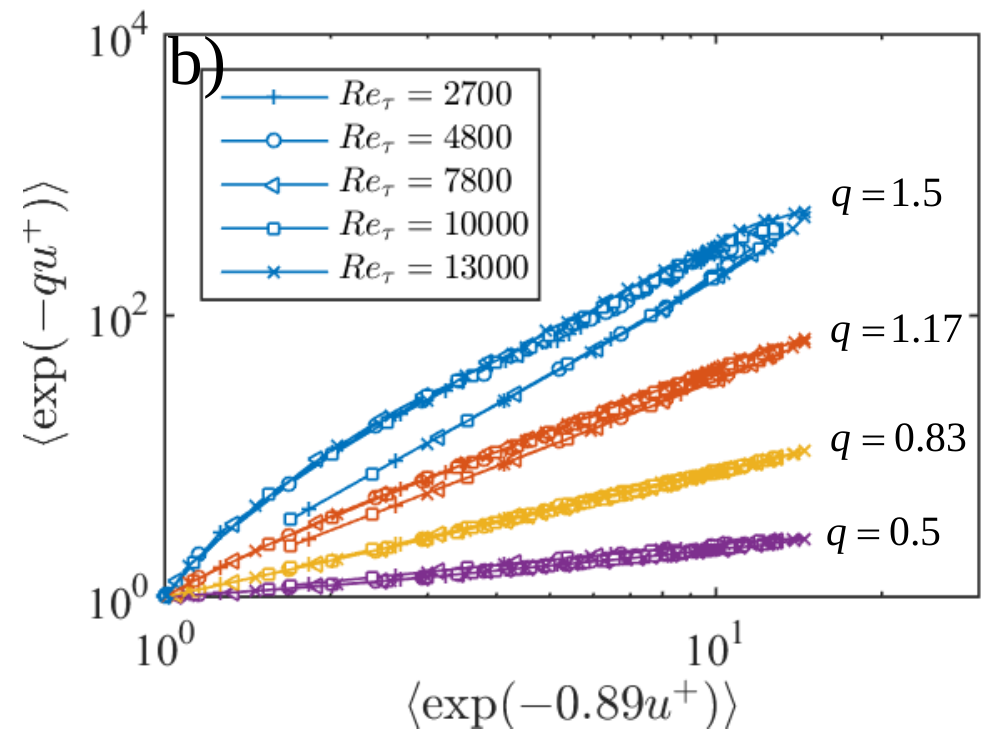}
\caption{Log-log plot of the measured $\left<\exp(qu^+)\right>$ against $\left<\exp(0.89u^+)\right>$ (a), b) respectively for $\left<\exp(qu^+)\right>$ and $\left<\exp(-qu^+)\right>$) for $q=0.5$ (purple), $0.83$ (yellow), $1.17$ (orange), $1.5$ (blue) at $Re_\tau=2700$ ($+$) , $4800$ ($\circ$), $7800$ ($\triangleleft$), $10000$ ($\Box$), $13000$ ($\times$). }
\label{fig:ESS-Re}
\end{figure}

The Reynolds number dependency is examined in Fig. \ref{fig:ESS-Re}. Hot-wire measurements of boundary layers at $Re_\tau=2700$, $4800$, $7800$, $10000$, $13000$ from the Melbourne HRNBLWT are analyzed. Details on this dataset can be found in Ref. \cite{marusic2015evolution}. 
In Fig. \ref{fig:ESS-Re}, the measured $W(\pm q;z/\delta)$ are plotted against the measured $W(\pm 0.89;z/\delta)$ for $Re_\tau=2700$, $4800$, $7800$, $10000$, $13000$ on a log-log scale. No significant Reynolds number dependency is found. 
Before concluding this section, let us take a look at the statistics for negative $q$ values. 
The region of extended self-similarity is narrower for $W(-q;z/\delta)$ compared to positive $q$-valued MGFs. As mentioned in this section, $\left<\exp(qu_z)\right>$ is dominated by fluctuations of the same sign as $q$, therefore $W(-q;z/\delta)$ emphasizes the ``ejection'' motions. For ``ejection'' motions, near wall fluid tends to be brought into the bulk region. Because viscosity is mostly dominant in the near wall region, footprints of viscous effects should be visible in such ``ejection'' motions. This is probably the reason why ESS scaling has a poorer quality compared to the cases where $q>0$ . Nevertheless, ESS shows much improved scaling properties compared to the standard scaling versus $\delta/z$ also for negative $q$ values (not shown). 

\section{Extended Self-similarity for Large-Scale and Small Scale Motions}\label{sect:ESS-uSL}

\begin{figure} 
\centering
\includegraphics[height=2.00in]{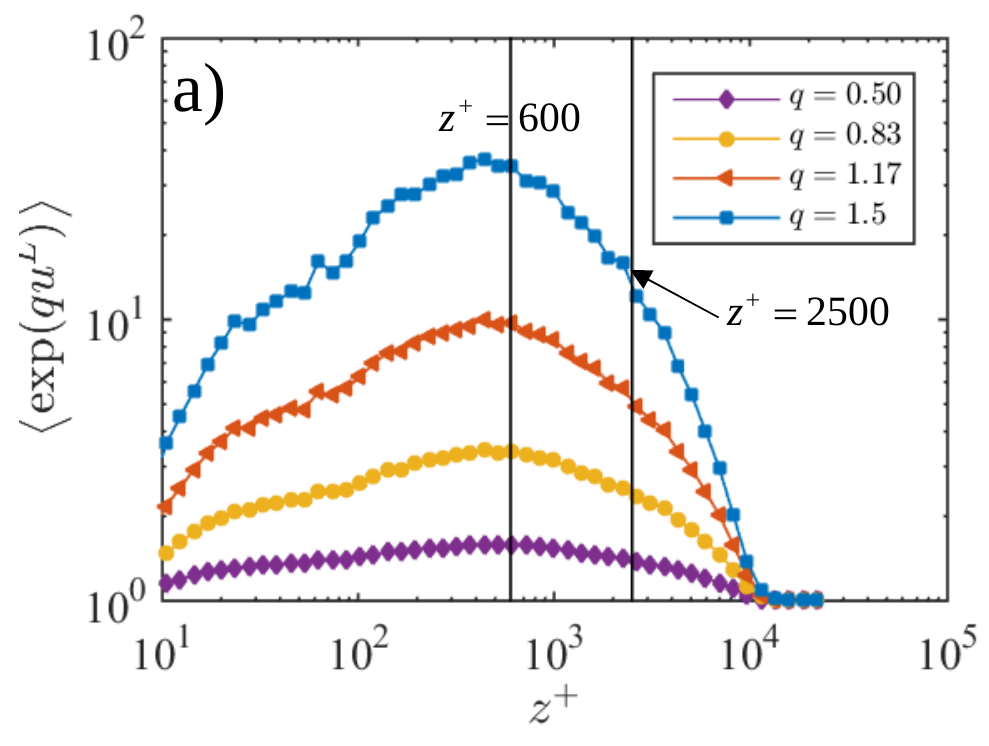}\\
\includegraphics[height=2.00in]{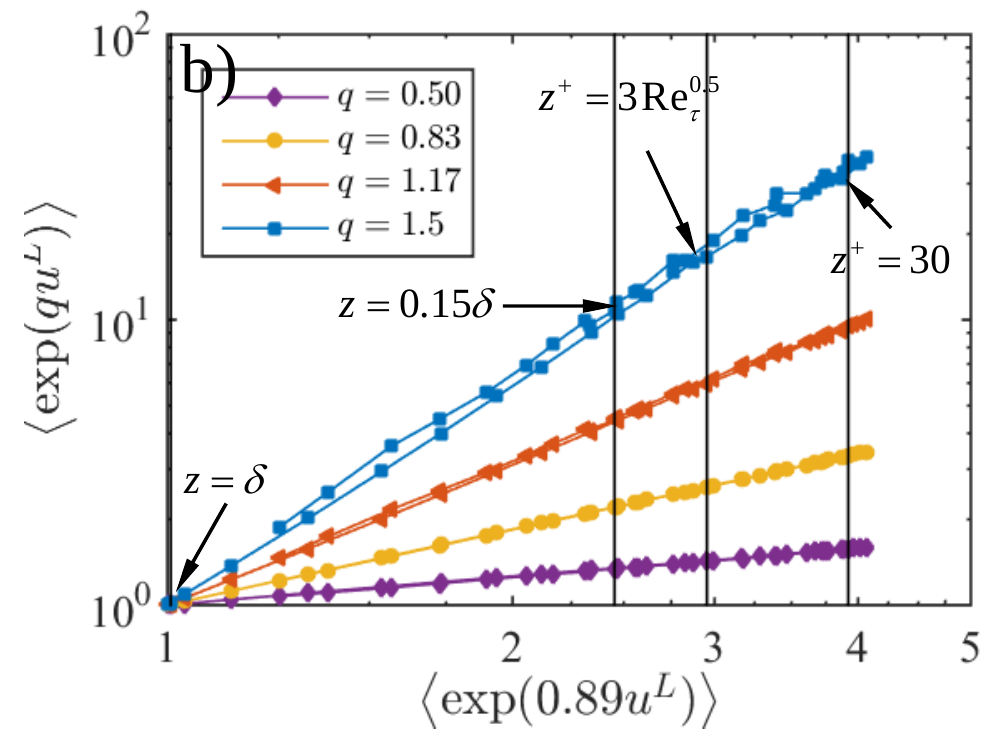}
\caption{a) log-log plot of $\left<\exp(qu^L_z)\right>$ against $z$ for $q=0.5$, $0.83$, $1.17$, $1.5$. The two vertical lines indicate $z^+=600$, $2500$. b) log-log plot of $\left<\exp(qu^L_z)\right>$ against $\left<\exp(0.89u^L_z)\right>$ for $q=0.5$, $0.83$, $1.17$, $1.5$. }
\label{fig:uL}
\end{figure}

\begin{figure} 
\centering
\includegraphics[height=2.00in]{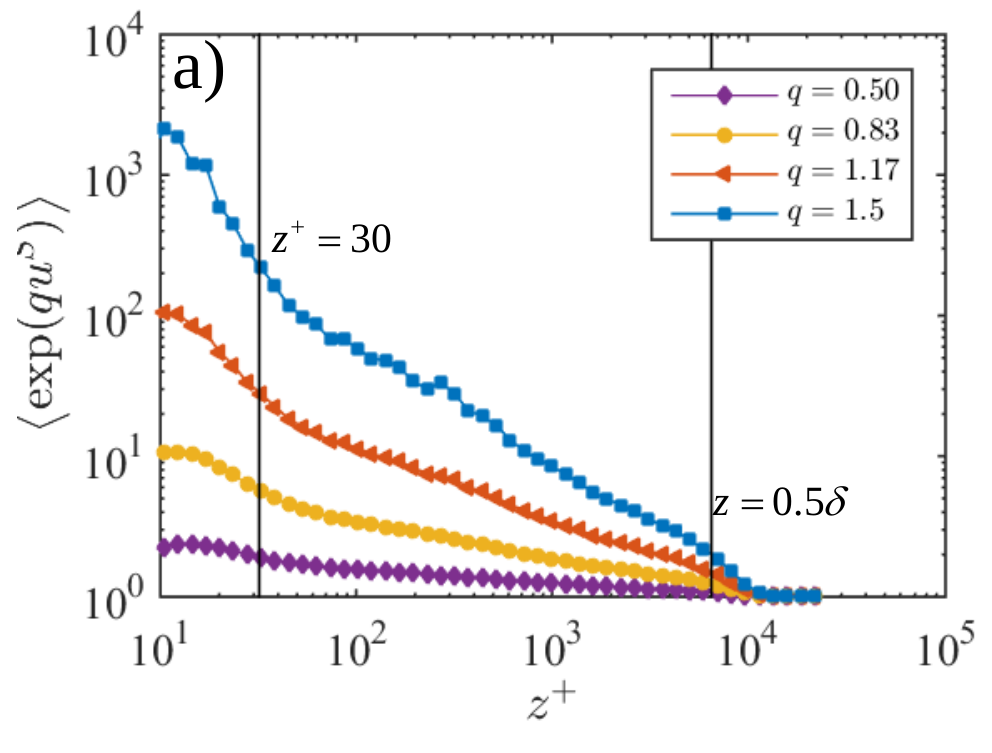}\\
\includegraphics[height=2.00in]{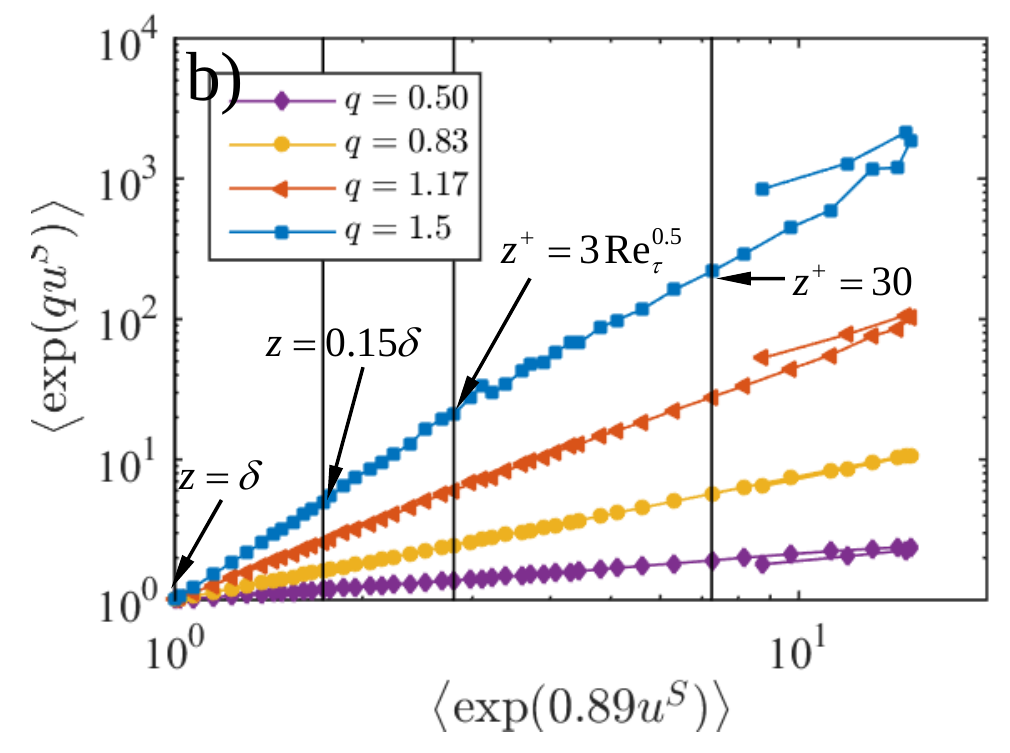}
\caption{Same as Fig. \ref{fig:uL} but for $u^S_z$. The two vertical lines indicates the wall normal heights $z^+=30$, $z=0.5\delta$. }
\label{fig:uS}
\end{figure}

In this section, empirical evidence of ESS scaling in large-scale and small scale fluid motions is presented. 
To obtain the velocity signal for large-scale motions, a top-hat filtering in Fourier space (using a cutoff wavenumber $k_\Delta=2\pi/\Delta$, where $\Delta$ is the filter scale) is conducted on the hot-wire measurements of a boundary layer at $Re_\tau=13000$. Boundary layer data at this Reynolds number provide us sufficient scale separation to decouple the viscosity-affected near-wall cycles and geometric-dependent large-scale motions \citep{mckeon2007asymptotic, hutchins2007evidence}.
The filtering length scale is one boundary layer height ($\Delta=\delta$) and is kept constant at all wall normal heights (50 measurement heights in total). The streamwise velocity fluctuations are thus decomposed into the large-scale, and small-scale fluctuations:
\begin{equation}
\small
u_z=u^L_z+u^S_z.
\end{equation}
$u^L_z$ is the filtered fluctuations, $u_z$ is the unfiltered velocity and $u^S_z=u_z-u^L_z$. 
The measured large-scale MGFs $\left<\exp(qu^L_z)\right>$ as functions of the wall distance are plotted in Fig. \ref{fig:uL}(a) for representative positive values of $q$. In Fig. \ref{fig:uL}(b), the same quantities are plotted in ESS scaling. In Fig. \ref{fig:uS}, the same plots for $u^S_z$ are shown. 
As one can see, power-law scaling of $\left<\exp(qu^L_z)\right>$ is barely seen in the log region. 
On the other hand, for $\left<\exp(qu^S_z)\right>$ (Fig. \ref{fig:uS} a)), power-law scaling is observed in the region $30< z^+$, $z<0.5\delta$. The ESS scalings are of high quality, both for $u^L_z$ and $u^S_z$. The region of ESS, from Figs. \ref{fig:uL} b), \ref{fig:uS} b), extends from $z^+=30$ to $z=\delta$.

\begin{figure} 
\centering
\includegraphics[height=2.00in]{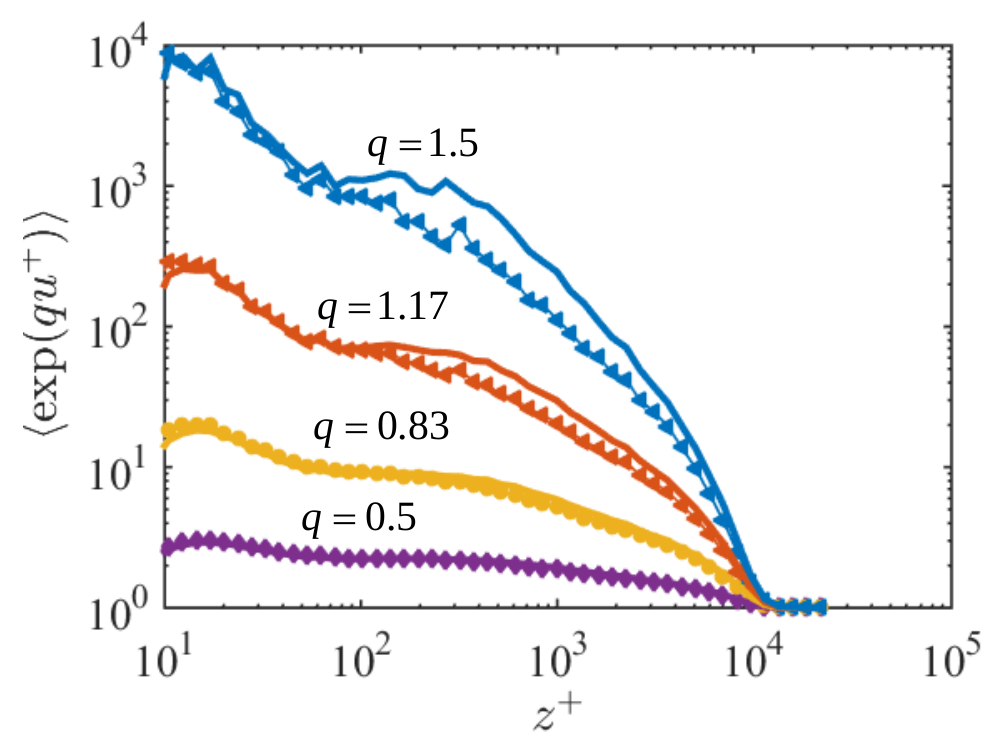}
\caption{Log-log plot of measured $\left<\exp(qu_z)\right>$ (symbols) against the wall normal distance $z$ for $q=0.5 (\diamond)$, $0.83 (\circ)$, $1.13 (\lhd)$. Lines are for $\left<\exp(qu_z^S)\right>\cdot\left<\exp(qu^L_z)\right>$: $q=0.5$ (purple), $0.83$ (yellow), $1.13$ (orange), $1.5$ (blue).
}
\label{fig:expqu-comp}
\end{figure}

From the above observations, we might conclude that the breaking of the power-law scaling of the MGFs is mainly brought in by the large-scale motions, which develop a sudden change in their statistical properties around $z^+ \approx 3 Re_\tau^{1/2}$. On the other hand, both the large-scale and small-scale components seem to exhibit good ESS scaling properties. 
As anticipated in \S \ref{sect:AttEddy}, we can rationalize these findings as follows: we model the velocity signal $u^S_z$ to be an additive process of i.i.d. Gaussian variables $a_i^S$ at all distances from $30 < z^+$ to $ z < \delta$, and $u^L_z$ to be another additive process of independent but not identically distributed Gaussian additives $a_i^L$ with the variance $\sigma_i^L$ depending on $z$:
\begin{equation}
\small
u^S_z=\sum_{i=1}^{N_z}a_i^S, ~~~u^L_z=\sum_{i=1}^{N_z}a_{i}^L.
\label{eq:uSL-rap}
\end{equation}
By invoking the independency among the random additives, one derives: 
\begin{equation}
\small
\begin{split}
\langle \exp(qu_z) \rangle 
&\sim 
\langle \prod_{i=1}^{N_z}\exp(qa_{i}^S)\rangle \langle \prod_{i=1}^{N_z}\exp(qa_{i}^L\rangle \\
&\sim 
\left(\frac{\delta}{z}\right)^{\tau^S(q)} 
\exp\left(\frac{q^2}{2}\sum_{i=1}^{N_z} (\sigma_i^L)\right). 
\end{split}
\label{eq:expquSL}
\end{equation}
with $\tau(q) = \sigma^S q^2/2$. We can then define the $z$-dependent variance $\sigma_i^L$ such as to reproduce the MGFs for $u^L_z$, $W^L(q;z) = \langle \exp(qu^L_z) \rangle$. To do that, we discretize the wall normal distance logarithmically: $z_i/\delta= 2^{-i}$ and we define $\sigma_i^L$ such that:
\begin{equation}
\small
\exp\left(\frac{q_o^2}{2}\sigma_i^L\right) = \frac{W^L(q_o,z_i)}{W^L(q_o,z_{i-1})}.
\label{eq:aiL}
\end{equation}
Replacing Eq. \ref{eq:aiL} in to Eq. \ref{eq:expquSL}, we obtain the ESS scaling: 
\begin{equation}
\small
\begin{split}
&\langle \exp(qu_z) \rangle \sim \left<\exp(q_ou_z)\right>^{(q/q_o)^2},\\
&\text{with}\; 
\left<\exp(q_ou_z)\right> =
\left(\frac{\delta}{z}\right)^{\sigma^S q_o^2/2} 
\frac{W^L(q_o;z/\delta)}{W^L(q_o;1)} 
\end{split}
\label{eq:ESSfinal}
\end{equation}
Hence ESS scaling is preserved in the range of scales where the two signals $u^L_z$ and $u^S_z$ can be represented by Eq. \ref{eq:uSL-rap}. In the above discussion, $u^S_z$, $u^L_z$ are considered statistically independent. For boundary layer flows, this cannot be exact (see e.g. Refs.\cite{marusic2010predictive, mathis2011predictive, ganapathisubramani2012amplitude}). 
To quantify possible deviations, in Fig. \ref{fig:expqu-comp} we compare $\left<\exp(qu_z)\right>$ against $\left<\exp(qu^L_z)\right>\cdot\left<\exp(qu^S_z)\right>$. The agreement between $\left<\exp(qu^L_z)\right>\cdot\left<\exp(qu^S_z)\right>$ and $\left<\exp(qu_z)\right>$ is in fact quite good (for $q\leq 1.17$). Because for high $q$ values $\left<\exp(qu)\right>$ emphasizes more rare, intense events, Fig. \ref{fig:expqu-comp} suggests correlation among large and small scale motions are mainly due to intense events. 

\section{Conclusions}\label{sect:conclusions}

Empirical evidence for ESS in the MGFs of the streamwise velocity fluctuations at high Reynolds number in the region $30<z^+$, $z<\delta$ is presented, indicating the existence of a common physical process in the log region, the bulk region and in the viscosity-affected region. Results are robust as a function of Reynolds number, at least in the range analyzed here. 
Relaxing the requirement of wall attached eddy being self-similar at all scales in the attached eddy model, we have shown that the hierarchical random additive processes can reproduce most observed scalings. Within the framework of the Townsend attached eddy hypothesis, present results suggest by allowing the eddy characteristic velocity scale to be dependent on the distance from the wall outside the log layer, the attached eddy model may be used to describe the flow beyond the log region. 
Then we split the velocity fluctuations in a filtered large-scale component, $u^L_z$ and a small-scale remaining component, $u^S_z=u_z-u^L_z$. A power-law scaling of $\left<\exp(qu^L_z)\right>$ is observed in the more restrictive log region, while for $\left<\exp(qu_z^S)\right>$ a more extended power-law scaling is observed in the region $100<z^+$, $z<0.5\delta$. ESS scaling is found over a wide range $30<z^+$, $z<\delta$ for both $u^L_z$ and $u^S_z$.
The effects of the filtering length scale and Reynolds number on the statistical properties of $u^L_z$, $u^S_z$ are left for future investigations, as well as the statistical structure of the seemingly self-similar process describing $u^S_z$. Studies of those effects can be helpful in understanding the interactions among the large and small scale motions in wall-bounded turbulence.

\section*{Acknowledgements}
The research leading to these results has received funding from the Office of Naval Research, the US National Science Foundation, the Australian Research Council, and the European Union's Seventh Framework Programme (FP7/2007-2013) under the ERC  grant agreement No. 339032. LB acknowledge useful discussions with R. Benzi and the kind hospitality of the Department of Mechanical Engineering at the Johns Hopkins University.




\bibliography{ESS_Yangetal16}

\end{document}